\documentclass[conference,a4paper]{IEEEtran}

\IEEEoverridecommandlockouts

\usepackage{graphicx} 
\usepackage{epstopdf} 

\usepackage{cite}
\usepackage{amsmath,amssymb,amsfonts}
\usepackage{algorithmic}
\usepackage{graphicx}
\usepackage{textcomp}
\usepackage{xcolor}
\usepackage{color}

\usepackage{amssymb}
\usepackage{bbm} 
\usepackage{bm}

\usepackage{algorithm}
\usepackage{algorithmic}
\usepackage{url}
\usepackage{color}

\usepackage{graphicx} 
\usepackage{epstopdf} 

\usepackage{cite}
\usepackage{amsmath,amssymb,amsfonts}
\usepackage{algorithmic}
\usepackage{graphicx}
\usepackage{textcomp}
\usepackage{xcolor}
\usepackage{color}

\usepackage{amssymb}
\usepackage{bbm} 
\usepackage{bm}

\usepackage{algorithm}
\usepackage{algorithmic}
\usepackage{url}
\usepackage{color}

\usepackage{multirow} 

\usepackage{subfig}

\usepackage{tablefootnote}

\usepackage{caption}

\usepackage{amsmath}
\allowdisplaybreaks[1]

\def\BibTeX{{\rm B\kern-.05em{\sc i\kern-.025em b}\kern-.08em
    T\kern-.1667em\lower.7ex\hbox{E}\kern-.125emX}}
\begin{document}

\title{Duration-Squeezing-Aware Communication and Computing for Proactive VR}
\author{\IEEEauthorblockN{Xing Wei, Chenyang Yang, and Shengqian Han}
\IEEEauthorblockA{School of Electronics and Information Engineering, Beihang University, Beijing 100191, China\\ Email: \{weixing, cyyang, sqhan\}@buaa.edu.cn}
}

\maketitle

\vspace{-18mm}
\begin{abstract}
    Proactive tile-based virtual reality video streaming computes and delivers the predicted tiles to be requested before playback. All existing works overlook the important fact that computing and communication (CC) tasks for a segment may squeeze the time for the tasks for the next segment,
    which will cause less and less available time for the latter segments.
    In this paper, 
    we jointly optimize the durations for CC tasks to maximize the completion rate of CC tasks under the task duration-squeezing-aware constraint. To ensure the latter segments remain enough time for the tasks, the CC tasks for a segment are not allowed to squeeze the time for computing and delivering the subsequent segment. We find the closed-form optimal solution, from which
    we find a minimum-resource-limited, an unconditional and a conditional resource-tradeoff regions, which are determined by the total time for proactive CC tasks and the playback duration of a segment. Owing to the duration-squeezing-prohibited constraints, the increase of the configured resources may not be always useful for improving the completion rate of CC tasks.
    Numerical results validate the impact of the duration-squeezing-prohibited constraints and illustrate the three regions.
\end{abstract}

\vspace{-2mm}\section{Introduction}
\vspace{-1mm}
Virtual reality (VR) video  requires 360$^{\circ}$$\times$ 180$^{\circ}$ panoramic view with ultra high resolution. Delivering such videos is cost-prohibitive for wireless networks. This inspires proactive tile-based streaming\cite{optimizing_VR,survey_Hsu},
which divides a full panoramic view segment into small tiles in spatial domain, predicts the future field of view (FoV) of a user, and then renders and transmits the tiles overlapped with the predicted FoVs.



Proactive tile-based VR video streaming contains three tasks: prediction, communication, and
computing. Given the predictor and the prediction accuracy required for satisfying the quality of experience (QoE), the total time for rendering and transmitting a segment can be determined\cite{Xing_VR_Shannon,verylong_predict,Fixation_Prediction}.
With such a total time budget, it has been shown in the literature that the communication and computing (CC) resources can be flexibly traded off \cite{VR_on_edge,millimeter_WiFi_VR}. For example, when the communication bandwidth is insufficient, one can assign more computing resource for rendering in order to provide longer time for delivering.

However, all existing works \cite{adaptive_TVT,MEC_THZ,wcnc_2020,ICME_2020,Xing_VR_Shannon} for proactive tile-based VR video streaming overlook an important fact: the communication and computing tasks for successive segments are coupled in timeline. Specifically, the communication task for multiple segments in a video forms a queue, and the computing task forms another queue. Transmitting and computing a segment may squeeze the time for tasks for the next segment, such that the QoE may degrade owing to the insufficient time left for accomplishing the tasks for latter segments.

In this paper, we investigate how to maximize the performance of proactive tile-based VR video streaming considering the coupled timeline for computing and delivering successive segments.
To this end, we jointly optimize the durations for these two tasks to maximize the completion rate of CC tasks under the duration-squeezing-aware constraint. When the length of a VR video is long, to ensure that the latter segments remain enough time for the tasks, the CC tasks for a segment are not allowed to squeeze
the time for computing and delivering the subsequent segment. We obtain the global optimal solution via Karush-Kuhn-Tucker (KKT) conditions. As far as the authors know, this is the first work that considers the time squeeze of these two tasks in proactive VR streaming.

From the closed-form solution of the optimal durations, we find a minimum-resource-limited, an unconditional and a conditional resource-tradeoff regions. The boundary of the three regions depends on the relative values of the total time budget for communication and computing as well as the playback duration of a segment.
In practice, these two durations can be very different, with the range of 0.2$\sim$10 seconds\cite{Predictive_xueshihou,verylong_predict,Fixation_Prediction} and the range of 0.5$\sim$2 seconds\cite{apcc,optimizing_VR,edge_assisted}, respectively, where the total time for proactive streaming highly depends on the predictor and the required prediction accuracy.
With different combinations of these two durations, the system may lie in one of the three regions.
In the minimum-resource-limited region, the communication and computing resources can not be traded off and the transmission and computing rates should be identical.
In the unconditional resource-tradeoff region, the resources can be flexibly traded off and increasing arbitrary resource is useful for improving the completion rate of CC tasks. In the conditional resource-tradeoff region, an extra condition should be satisfied to achieve the tradeoff and improving the completion rate of CC tasks.

\vspace{-0.2cm}
\section{System Model}\label{section-system_model}
\vspace{-1mm}
Consider a proactive tile-based VR video streaming system with a mobile edge computing (MEC) server co-located with a base station (BS).
Each VR video consists of $L$ segments in temporal domain, and each segment consists of $M$ tiles in spatial domain. The playback duration of each tile equals to the playback duration of a segment, denoted by $T_{\mathrm{seg}}$\cite{optimizing_VR,survey_Hsu}.
Each user is equipped with a head-mounted display (HMD), which can measure the head movement data, send the data to the MEC server, and pre-buffer segments.
The MEC server renders a video segment before delivering to the HMD.



\begin{figure}[htbp]
    \centering
    \subfloat[Rendering and transmitting pipeline squeeze, $\Delta p \!>0,\Delta m \!>0$]{\label{Fig:MEC4VR_pipeline-TTC_1}
        \begin{minipage}[c]{0.7\linewidth}
            \centering
            \includegraphics[width=1\textwidth]{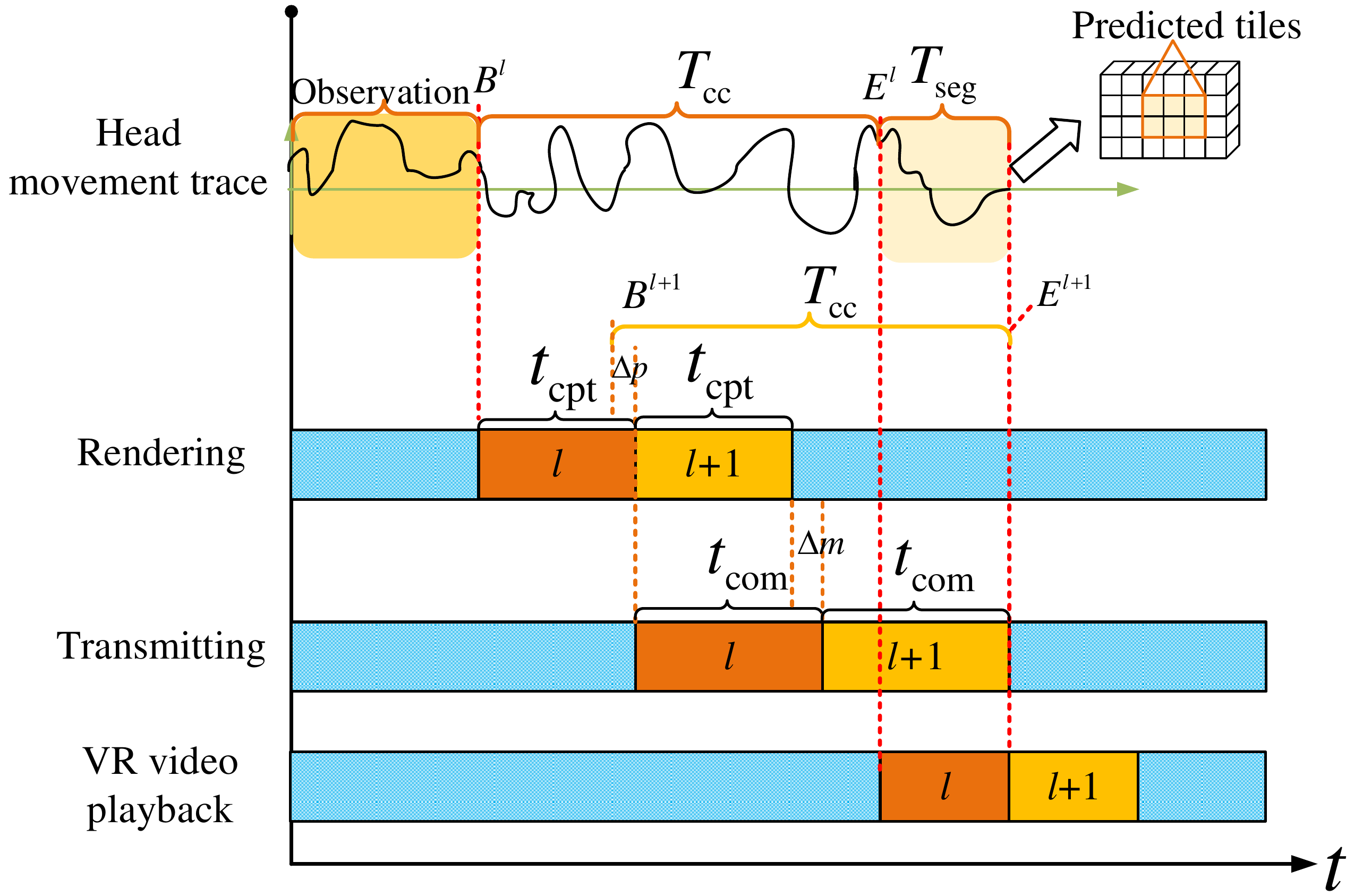}
        \end{minipage}
    }\\ \vspace{-2mm}
    \centering
    \subfloat[Transmitting pipeline squeeze, $\Delta p <0,\Delta m >0$]{\label{Fig:MEC4VR_pipeline-TTC_2}
        \begin{minipage}[c]{0.7\linewidth}
            \centering
            \includegraphics[width=1\textwidth]{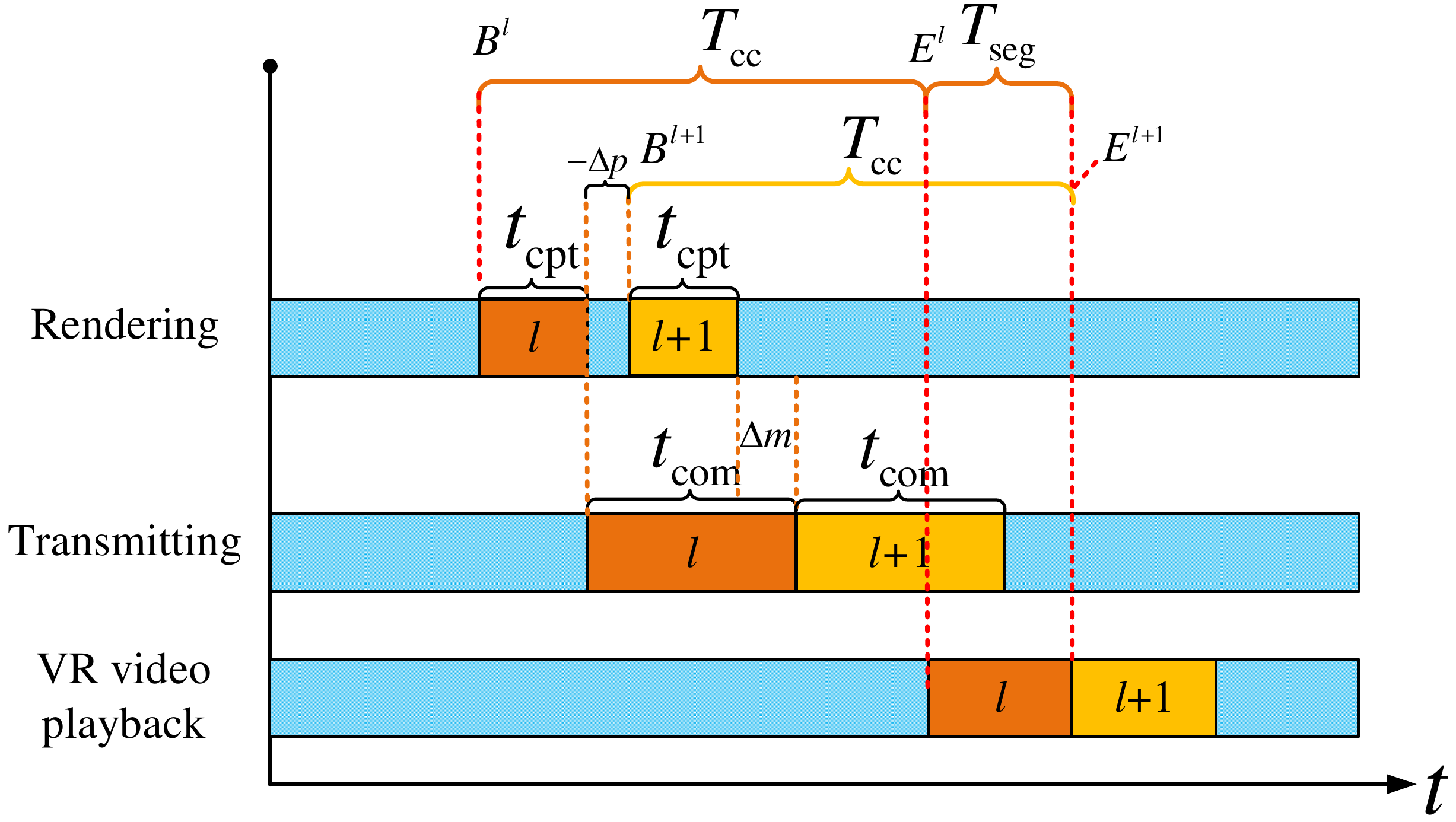}
        \end{minipage}
    }
    \vspace{-0.1cm}
    \caption{Proactively streaming the $l$th and ($l$+1)th segments.}\label{Fig:MEC4VR_pipeline-TTC}	
    \vspace{-0.2cm}
\end{figure}

When a user requests a VR video, the MEC server first streams the first $l-1$ segments in a reactive or a passive mode\cite{transmission_mode-standard-update}. When the MEC server collects the information of the user (e.g., the head movement data) in an observation window, proactive streaming for the $l$th segment begins, then subsequent segments are predicted, rendered, and transmitted one after another, as shown in Fig. 
\ref{Fig:MEC4VR_pipeline-TTC_1}.
Specifically, at the end of the observation window for the $l$th segment, i.e, $B^l$, the tiles in the $l$th segment to be requested 
are first predicted, 
then the predicted tiles are rendered with duration $t_{\mathrm{cpt}}$, and finally the rendered tiles are transmitted with duration $t_{\mathrm{com}}$, which should be finished before the start time for playback for the segment, i.e., $E^l$. Therefore, the total computing and transmission time for the segment $T_{\mathrm{cc}}=E^l - B^l$.

To train a predictor for the whole video, $T_{\mathrm{cc}}$ for every segment needs to be identical. A predictor can be more accurate with a smaller value of $T_{\mathrm{cc}}$. This is because the tiles to be predicted are closer to and hence are more correlated with the head movement sequence in the observation window\cite{Xing_VR_Shannon}. Given a predictor and required viewport prediction accuracy,
the value of $T_{\mathrm{cc}}$ can be determined\cite{Xing_VR_Shannon,verylong_predict,Fixation_Prediction}.



\vspace{-0.1cm}
\subsection{Duration-Squeezing-Aware Constraint}
\vspace{-0.05cm}
With identical value of $T_{\mathrm{cc}}$ for every segment, we can observe that $E^{l+1}-E^l=B^{l+1} - B^l$. Without playback stalling, $E^{l+1}-E^l=T_{\mathrm{seg}}$ holds and thus $B^{l+1}-B^l=T_{\mathrm{seg}}$. If the rendering for the $l$th segment finishes after $B^{l+1}$, then the computing task will squeeze the time for rendering the ($l$+1)th segment. Denote the squeezed computing time as $\Delta p=t_{\mathrm{cpt}} - (B^{l+1} - B^l)=t_{\mathrm{cpt}} - T_{\mathrm{seg}}$.
If the rendering for the $l$th segment can be finished within $T_{\mathrm{seg}}$, then $\Delta p\leq0$, and there is no squeeze in rendering, as shown in Fig. \ref{Fig:MEC4VR_pipeline-TTC_2}.

Similarly, the communication task may also squeeze the time for delivering the ($l$+1)th segment. Denote the squeezed communication time as $\Delta m$. When $\Delta p>0$, $\Delta m = t_{\mathrm{com}} - t_{\mathrm{cpt}}$, as shown in Fig. \ref{Fig:MEC4VR_pipeline-TTC_1}. When $\Delta p<0$, $\Delta m = t_{\mathrm{com}} - t_{\mathrm{cpt}} - (-\Delta p)$, as shown in Fig. \ref{Fig:MEC4VR_pipeline-TTC_2}. By summarizing the two cases, we obtain $\Delta m=t_{\mathrm{com}} - t_{\mathrm{cpt}} - (-\Delta p)^{+}$,
where $(x)^+\triangleq\max\{x,0\}$. When $\Delta m \leq 0$, the transmission can be finished on time and there is no squeeze in the pipeline.

For the $l$th segment, which is the first segment with proactive streaming, the transmission and rendering tasks should be finished within $T_{\mathrm{cc}}$, i.e., $t_{\mathrm{cpt}} + t_{\mathrm{com}} \leq T_{\mathrm{cc}}$.
For the ($l$+1)th segment, the remaining duration for CC tasks is $t_{\mathrm{cpt}}  + t_{\mathrm{com}}\leq T_{\mathrm{cc}} - \left((\Delta p)^+ + (\Delta m)^+\right)$.
For the $L$th segment, the remaining duration for CC tasks is
\begin{align}\label{duration_final}
t_{\mathrm{cpt}}  + t_{\mathrm{com}} \leq T_{\mathrm{cc}}  - (L - l)\left\{(\Delta p)^+ + (\Delta m)^+\right\}.
\end{align}
\vspace{-0.1cm}
\subsection{Computing and Transmission Model}
\vspace{-0.05cm}
The computing resource of MEC for rendering a VR video can be assigned by allocating graphics processing unit (GPU) and compute unified device architecture cores \cite{Nvidia_cloudXR,Xing_VR_Shannon}.
To gain useful insight, we assume that the computing resource, denoted as $\mathcal{C}_{\mathrm{total}}$ (in floating-point operations per second, FLOPS), is equally allocated among $K$ users. Then,  the number of bits that can be rendered per second, referred to as the \textit{computing rate}, for the $k$th user, is
\begin{align*}
C_{\mathrm{cpt},k} \triangleq \frac{\mathcal{C}_{\mathrm{total}}}{K\cdot\mu_r} (\textit{in bit/s}),
\end{align*}
where $\mu_r$ is the required floating-point operations (FLOPs) for rendering one bit of FoV \textit{in FLOP/bit}\cite{Xing_VR_Shannon}.

The BS serves $K$ single-antenna users using zero-forcing beamforming over bandwidth $B$ with $N_t$ antennas.
The instantaneous data rate at the $i$th time slot for the $k$th user is
\begin{align*}
C_{\mathrm{com},k}^{i}=B\log_2 \left(1+\frac{p_k d_k^{-\alpha}|\tilde{h}^i_k|^2}{\sigma^2} \right),
\end{align*}
where $\tilde{h}^i_k\triangleq (\mathbf{h}^i_k)^H\mathbf{w}^i_k$ is the equivalent channel gain, $p_k$ and $\mathbf{w}^i_k$ are respectively the transmit power and beamforming vector for the $k$th user,  $d_k$ and $\mathbf{h}^i_k\in\mathbb{C}^{N_t}$ are respectively the distance and the small scale channel vector from the BS to the $k$th user, $\alpha$ is the path-loss exponent, $\sigma^2$ is the noise power, and $(\cdot)^{H}$ denotes conjugate transpose.

We consider indoor users as in the literature, where the distances of users, $d_k$, usually change slightly \cite{LSTM_update,survey_Hsu,NTHU_dataset} and hence are assumed fixed.
Due to the head movement and the variation of the environment,
small-scale channels are time-varying, which are assumed as remaining constant in each time slot with duration $\Delta T$ and changing independently with identical distribution among time slots.
With the proactive transmission, the predicted FoVs in a segment should be transmitted with duration $t_{\mathrm{com}}$. The number of bits transmitted with $t_{\mathrm{com}}$
can be expressed as $\overline{C}_{\mathrm{com},k}t_{\mathrm{com}}$,
where $\overline{C}_{\mathrm{com},k} \triangleq \frac{1}{N_s}\sum_{i=1}^{N_s}C_{\mathrm{com},k}^{i} \Delta T$ is the time average transmission rate,
and $N_s$ is the number of time slots in $t_{\mathrm{com}}$.
Since future channels are unknown when optimizing the durations, we use ensemble-average rate $\mathbbm{E}_{\mathbf{h}}\{C_{\mathrm{com},k}^i\}$ to approximate the time-average rate $\overline{C}_{\mathrm{com},k}$,
which is very accurate when $N_s$ or $N_t/K$ is large \cite{Xing_VR_Shannon}.
To ensure fairness among users in terms of QoE, the transmit power is used to compensate for the path loss, i.e., ${p}_k=\frac{\beta}{d_k^{-\alpha}}$, where $\beta$ can be obtained from $\beta(\sum_{k=1}^{K}\frac{1}{d_k^{-\alpha}})={P}$ and ${P}$ is the maximal transmit power of the BS. Then, the ensemble-average transmission rate for each user is equal.

Without loss of generality, we consider an arbitrary user for analysis in the sequel. For notational simplicity, we use $C_{\mathrm{com}}$ to represent $\mathbbm{E}_{\mathbf{h}}\{C_{\mathrm{com},k}^{i}\}$ and use $C_{\mathrm{cpt}}$ to represent $C_{{\mathrm{cpt}},k}$.

\vspace{-1.5mm}
\section{Duration Optimization for Computing and Communication}\label{three_regions}
\vspace{-1.2mm}

To reflect the system performance for rendering and delivering all the predicted FoVs in a segment, define the completion rate of communication and computing (CC) tasks as 
\begin{align}
  S_{\mathrm{cc}} \triangleq \min\left\{\frac{C_{\mathrm{com}}t_{\mathrm{com}}}{S_{\mathrm{com}}^{}}, \frac{C_{\mathrm{cpt}}t_{\mathrm{cpt}}}{S_{\mathrm{cpt}}^{}}\right\},\label{S-cc}
  \vspace{-1.0mm}
\end{align}
where $S_{\mathrm{com}} = s_{\textit{fov}}\cdot r_f \cdot T_{\mathrm{seg}}/\gamma_c$ and $S_{\mathrm{cpt}} = s_{\textit{fov}}\cdot r_f \cdot T_{\mathrm{seg}}$
are respectively the number of bits of all the predicted FoVs in a segment for transmission \cite{HuaWei_Cloud_VR} and for rendering,
$\gamma_c$ is  the video compression ratio, $r_f$ (in frames per second) is frame rate,  $s_{\textit{fov}}\triangleq\gamma_{\textit{fov}} R_w R_h b$ is the number of bits in a FoV,
$\gamma_{\textit{fov}}$ is the ratio of FoV in a frame,
$R_w$ and $R_h$ are respectively the pixels in wide and high of a frame, and $b$ is the number of bits per pixel relevant to color depth\cite{HuaWei_Cloud_VR}. 
By substituting $S_{\mathrm{com}}$  and  $S_{\mathrm{cpt}}$ into \eqref{S-cc}, we obtain
\begin{align}
S_{\mathrm{cc}}= \frac{\min\{\tilde{C}_{\mathrm{com}}t_{\mathrm{com}}, C_{\mathrm{cpt}}t_{\mathrm{cpt}}\}}{s_{\textit{fov}}\cdot r_f \cdot T_{\mathrm{seg}}}, \vspace{-1.0mm}\label{S-cc-trans}
\end{align}
where $\tilde{C}_{\mathrm{com}}\triangleq C_{\mathrm{com}}\gamma_c$ is the equivalent transmission rate.

If $S_{\mathrm{cc}}>$ 100\%, the system is capable of streaming more bits beyond all the predicted FoVs. The extra capability can be used to stream the tiles at the marginal region of the predicted viewport\cite{FoV_edge_streaming,HuaWei_Cloud_VR} to compensate for the prediction errors and increase
the overlap of the delivered tiles and the requested
tiles in a segment.
If $S_{\mathrm{cc}}=0$, the HMD cannot receive any rendered FoV on time, which will cause playout stalls.


The durations for computing and delivering are optimized to maximize the completion rate of CC tasks, i.e.,
\vspace{-0.1cm}
\begin{subequations}
\label{P0}
\begin{align}
\textbf{P0}: & \max_{t_{\mathrm{cpt}}, t_{\mathrm{com}}} \  \ \ \ \ \ S_{\mathrm{cc}} \label{P1_pre_obj}\\
&  \ \ \ \  s.t.    \ \ \ \ \ \Delta p=t_{\mathrm{cpt}} - T_{\mathrm{seg}}, \label{squeeze_t_cpt_3}\\
& \ \ \ \ \ \ \ \    \ \ \ \ \ \Delta m=t_{\mathrm{com}} - t_{\mathrm{cpt}} - (-\Delta p)^{+},\label{squeeze_t_com_3} \\
& t_{\mathrm{cpt}}  + t_{\mathrm{com}} \leq T_{\mathrm{cc}} - (L - l)\left\{(\Delta p)^+ + (\Delta m)^+\right\}. \label{duration_final_3}
\end{align}
\end{subequations}

Problem \textbf{P0} contains four cases, depends on whether or not $\Delta p$ and $\Delta m$ exceed zero.
When the length of a VR video (i.e., $L$) is long, to ensure that every latter segment has time to be rendered and delivered, i.e.,
the right-hand side of \eqref{duration_final_3} is larger than zero, the values of $\Delta p$ and $\Delta m$ should be non-positive. That is to say,
squeezing  either transmission or rendering time of the subsequent segment is strictly prohibited.
When $\Delta p \leq 0$, we obtain $t_{\mathrm{cpt}}\leq T_{\mathrm{seg}}$  from \eqref{squeeze_t_cpt_3}.
When $\Delta m \leq 0$, by substituting \eqref{squeeze_t_cpt_3} into \eqref{squeeze_t_com_3}, we obtain $t_{\mathrm{com}}\leq T_{\mathrm{seg}}$.
Then, problem \textbf{P0} degenerates into
\vspace{-0.1cm}
\begin{subequations}
\label{P1-all}
\begin{align}
  \textbf{P1}:& \max_{t_{\mathrm{cpt}}, t_{\mathrm{com}}} \  \ \ \ \ \ S_{\mathrm{cc}} \label{P1_obj}\\
  &  \ \ \ \  s.t.    \ \ \ \ \ t_{\mathrm{cpt}}+t_{\mathrm{com}}\leq T_{\mathrm{cc}},\label{T_cc_def}\\
 & \ \ \ \ \  \  \ \  \ \ \ \ \ t_{\mathrm{cpt}}\leq T_{\mathrm{seg}},\label{queue_stable_cpt}\\
  & \ \ \ \ \  \  \ \  \ \ \ \ \ t_{\mathrm{com}}\leq T_{\mathrm{seg}}.\label{queue_stable_com}
\end{align}
\end{subequations}
Problem \textbf{P1} can be transformed into a convex problem. From the KKT conditions, its optimal solution and the maximal value of the objective function of \textbf{P1} can be obtained as
\begin{subequations}\label{opt_slt}
\begin{align}
&t_{\mathrm{cpt}}^*
\left
\{\begin{array}{lr}
\in\left[\frac{\tilde{C}_{\mathrm{com}}T_{\mathrm{seg}}}{C_{\mathrm{cpt}}},T_{\min}\right],& \tilde{C}_{\mathrm{com}}< C_{\mathrm{cpt}}\ \textrm{and} \ T_{\mathrm{c}}^{\max}> T_{\mathrm{seg}}, \\
=T_{\mathrm{seg}},& \tilde{C}_{\mathrm{com}}\geq C_{\mathrm{cpt}} \ \textrm{and}  \ T_{\mathrm{c}}^{\max}> T_{\mathrm{seg}}, \\
=\frac{\tilde{C}_{\mathrm{com}}T_{\mathrm{cc}}}{\tilde{C}_{\mathrm{com}} + C_{\mathrm{cpt}}}, & T_{\mathrm{c}}^{\max}\leq T_{\mathrm{seg}},
\end{array}
\right.\label{opt_slt_t_cpt}\\
&t_{\mathrm{com}}^*
\left
\{\begin{array}{lr}
=T_{\mathrm{seg}},& \tilde{C}_{\mathrm{com}}\leq C_{\mathrm{cpt}}\ \textrm{and}  \ T_{\mathrm{c}}^{\max}> T_{\mathrm{seg}}, \\
\in\left[\frac{C_{\mathrm{cpt}}T_{\mathrm{seg}}}{\tilde{C}_{\mathrm{com}}},T_{\min}\right],& \tilde{C}_{\mathrm{com}}> C_{\mathrm{cpt}} \ \textrm{and}  \   T_{\mathrm{c}}^{\max}> T_{\mathrm{seg}}, \\
=\frac{C_{\mathrm{cpt}}T_{\mathrm{cc}}}{\tilde{C}_{\mathrm{com}} + C_{\mathrm{cpt}}}, & T_{\mathrm{c}}^{\max}\leq T_{\mathrm{seg}},
\end{array}
\right.\label{opt_slt_t_com}\\
&S_{\mathrm{cc}}^* =
\left
\{\begin{array}{lr}
\frac{\min\{\tilde{C}_{\mathrm{com}},C_{\mathrm{cpt}}\}}{s_{\textit{fov}}\cdot r_f},&\ T_{\mathrm{c}}^{\max}> T_{\mathrm{seg}}, \\
\frac{\tilde{C}_{\mathrm{com}}C_{\mathrm{cpt}}T_{\mathrm{cc}}}{s_{\textit{fov}}\cdot r_f\cdot T_{\mathrm{seg}}(\tilde{C}_{\mathrm{com}} + C_{\mathrm{cpt}})}, & T_{\mathrm{c}}^{\max}\leq T_{\mathrm{seg}},
\end{array}
\right.\label{opt_slt_S_cc}
\end{align}
\end{subequations}
where $T_{\min}\triangleq \min\{T_{\mathrm{cc}} - T_{\mathrm{seg}}, T_{\mathrm{seg}}\}$ and
\begin{align}\label{T_c_max}
T_{\mathrm{c}}^{\max}\triangleq\frac{\max\{\tilde{C}_{\mathrm{com}},C_{\mathrm{cpt}}\}T_{\mathrm{cc}}}{\tilde{C}_{\mathrm{com}} + C_{\mathrm{cpt}}} = \max\{t^o_{\mathrm{cpt}},t^o_{\mathrm{com}}\}.\vspace{-4.0mm}
\end{align}
$t^o_{\mathrm{cpt}}$ and $t^o_{\mathrm{com}}$ are the optimal durations for computing and communication without the constraints in \eqref{queue_stable_cpt} and \eqref{queue_stable_com} as  considered in \cite{Xing_VR_Shannon}.

\vspace{-1.5mm}
\section{Minimum-Resource-Limited, Unconditional and Conditional Resource-Tradeoff Regions}
\vspace{-1.5mm}
In this section, we show that the system may operate in a minimum-resource-limited, an unconditional resource-tradeoff, or a conditional resource-tradeoff regions.

First we discuss the two cases in \eqref{opt_slt_S_cc}.

\textit{Case 1} $T_{\mathrm{c}}^{\max} > T_{\mathrm{seg}}$: If $\tilde{C}_{\mathrm{com}}>C_{\mathrm{cpt}}$, then $T_{\mathrm{c}}^{\max}=t^o_{\mathrm{cpt}}$ from \eqref{T_c_max}.
Since the allowed maximal duration for rendering is $T_{\mathrm{seg}}$ as shown in \eqref{queue_stable_cpt}, $T_{\mathrm{c}}^{\max} > T_{\mathrm{seg}}$ indicates that $t^o_{\mathrm{cpt}}$ exceeds the allowed rendering duration. This suggests that the completion rate of CC tasks is limited by the computing rate,
where increasing the other type of resource $\tilde{C}_{\mathrm{com}}$ is useless for improving the system performance. Similarly, if $\tilde{C}_{\mathrm{com}}<C_{\mathrm{cpt}}$, then $T_{\mathrm{c}}^{\max}=t^o_{\mathrm{com}}$ and the system performance is limited by the transmission rate. 
We refer to this case as ``Minimum-resource-limited case",
where the efficient resource configuration should satisfy $\tilde{C}_{\mathrm{com}}=C_{\mathrm{cpt}}$.

We refer to a resource configuration as ``\textit{efficient}" when
the decrease of arbitrary one type of resources in the configuration will reduce the value of $S_{\mathrm{cc}}^* $.

\textit{Case 2} $T_{\mathrm{c}}^{\max}\leq T_{\mathrm{seg}}$: Both $t^o_{\mathrm{cpt}}$ and $t^o_{\mathrm{com}}$ satisfy the duration-squeezing-prohibited constraints in \eqref{queue_stable_cpt} and \eqref{queue_stable_com}. In this case, either increasing the computing rate or the transmission rate can improve the completion rate of CC tasks. 
This indicates a tradeoff between the computing rate and transmission rate \cite{Xing_VR_Shannon}.  %
We refer to this case as ``Resource-tradeoff case", where the resource configuration is flexible.

However, the boundary of the two cases depends on $T_{\mathrm{c}}^{\max}$,
which further depends on $\tilde{C}_{\mathrm{com}}$ and $C_{\mathrm{cpt}}$ as shown in \eqref{T_c_max}.
To provide useful insight into the resource configuration,
we provide three regions in the following, which are independent of the configured resources.


According to \eqref{T_c_max}, we have
\begin{align}\label{region_inequality}
T_{\mathrm{cc}}> T_{\mathrm{c}}^{\max} \geq \frac{T_{\mathrm{cc}}}{2}.
\end{align}

\textit{Minimum-resource-limited region}: If $\frac{T_{\mathrm{cc}}}{2} > T_{\mathrm{seg}}$, then with $T_{\mathrm{c}}^{\max}\geq \frac{T_{\mathrm{cc}}}{2}$ we have $T_{\mathrm{c}}^{\max}> T_{\mathrm{seg}}$, i.e., \emph{Case 1} holds.

\textit{Unconditional resource-tradeoff region}: If $T_{\mathrm{cc}}\leq T_{\mathrm{seg}}$, then with $T_{\mathrm{c}}^{\max}< T_{\mathrm{cc}}$ we have $T_{\mathrm{c}}^{\max}< T_{\mathrm{seg}}$, which is the sufficient condition to make \emph{Case 2} satisfied.

\textit{Conditional resource-tradeoff region}: If $T_{\mathrm{cc}}\in(T_{\mathrm{seg}},2T_{\mathrm{seg}}]$,
considering that $\frac{\max\{C_{\mathrm{com}},C_{\mathrm{cpt}}\}}{C_{\mathrm{com}} + C_{\mathrm{cpt}}}\in[\frac{1}{2},1)$, we obtain $T_{\mathrm{c}}^{\max} \in(\frac{T_{\mathrm{seg}}}{2}, 2T_{\mathrm{seg}})$. The system may operate in \emph{Case 1} or \emph{Case 2}.
If $T_{\mathrm{c}}^{\max}\leq T_{\mathrm{seg}}$, then the system lies in \emph{Case 2}. If $T_{\mathrm{c}}^{\max}> T_{\mathrm{seg}}$, then the system lies in \emph{Case 1}, where the efficient resource configuration is $C_{\mathrm{com}}=C_{\mathrm{cpt}}$
and we have $T_{\mathrm{c}}^{\max}= \frac{T_{\mathrm{cc}}}{2}$  from \eqref{T_c_max}. Further considering one boundary of the region $T_{\mathrm{cc}}\leq 2T_{\mathrm{seg}}$, we obtain $T_{\mathrm{c}}^{\max}\leq T_{\mathrm{seg}}$, which is the condition of \emph{Case 2} and can also be re-written as the condition for the efficient resource configuration as $\frac{\max\{C_{\mathrm{com}},C_{\mathrm{cpt}}\}}{C_{\mathrm{com}} + C_{\mathrm{cpt}}}\leq \frac{T_{\mathrm{seg}}}{T_{\mathrm{cc}}}$.
That is to say, in this region {even if in \emph{Case 1}, the efficient resource configuration can transform the system into \emph{Case 2}, i.e., the resource-tradeoff case}.

\vspace{-0.1cm}\section{Numerical Results}\label{numerical_results}\vspace{-0.1cm}
In this section, we validate the obtained analytical results and evaluate the performance of the optimized durations.

We consider the VR video with 4K resolution (3840$\times$2160 pixels\cite{FoV_aware_tile}) and $b=12$ bits per pixel\cite{HuaWei_Cloud_VR}. The ratio of a FoV to a frame is $\gamma_{\textit{fov}}=0.2$\cite{NTHU_dataset}, then the number of bits in a FoV is $s_\textit{fov}=3840\times2160\times b\times \gamma_{\textit{fov}}=19.9$ Mbits. The frame rate of VR video is $r_f=30$ frames per second \cite{FoV_aware_tile}. The compression ratio is  $\gamma_c=2.41$\cite{HEVC_lossless_coding}.  The playback duration of a segment is $T_{\mathrm{seg}}=1$~s\cite{FoV_aware_tile}. Depending on the configured communication and computing resources as well as the number of users, the computing and transmission rates for a user can be very different. For example, when $K=4$, $N_t=8$, $P=24$ dBm, $B=40$ MHz, and $d_k=5$ m, the ensemble-average transmission rate for a user is $C_{\mathrm{com}}=0.78$ Gbps\cite{Xing_VR_Shannon}, and the equivalent transmission rate $\tilde{C}_{\mathrm{com}}=C_{\mathrm{com}}\gamma_c=1.87$ Gbps.  When Nvidia P40 GPU is used for rendering VR videos for four users, the computing rate for a user is $C_{\mathrm{cpt}}=1.6$ Gbps\cite{Xing_VR_Shannon}. To reflect the variation of configured resources, we set $\tilde{C}_{\mathrm{com}}, C_{\mathrm{cpt}}\in[0,1]$ Gbps, unless otherwise specified.

\begin{figure}[htbp]
\vspace{-0.5cm}
    \centering
    \subfloat[$T_{\mathrm{cc}}>2T_{\mathrm{seg}} (T_{\mathrm{cc}}=2.1$ s)]{\label{Fig:MinResReg}
        \begin{minipage}[c]{0.5\linewidth}
            \centering
            \includegraphics[width=1\textwidth]{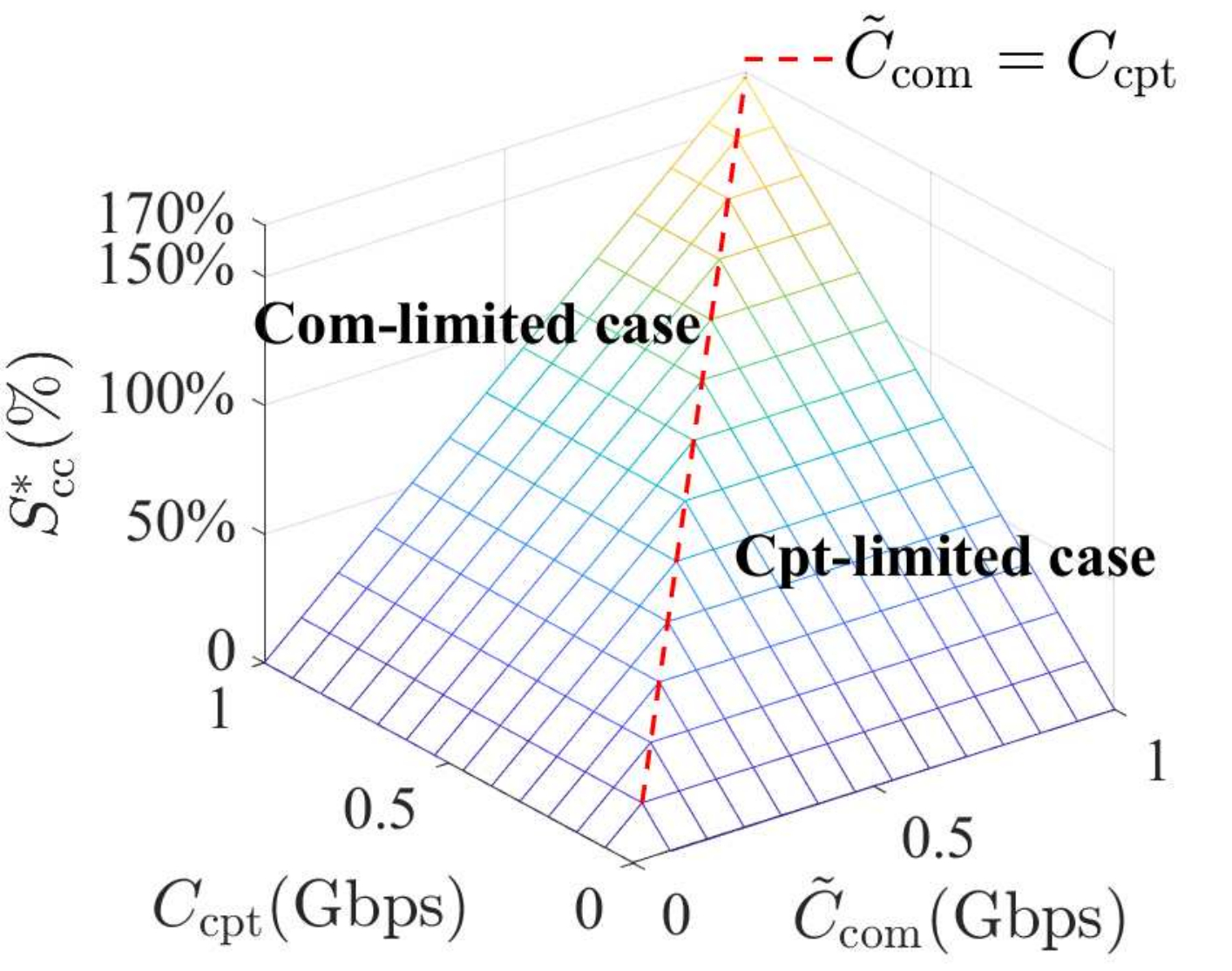}
        \end{minipage}
    }
    \subfloat[$T_{\mathrm{cc}}< T_{\mathrm{seg}} (T_{\mathrm{cc}}=0.9$ s)]{\label{Fig:ResMatReg}
        \begin{minipage}[c]{0.5\linewidth}
            \centering
            \includegraphics[width=1\textwidth]{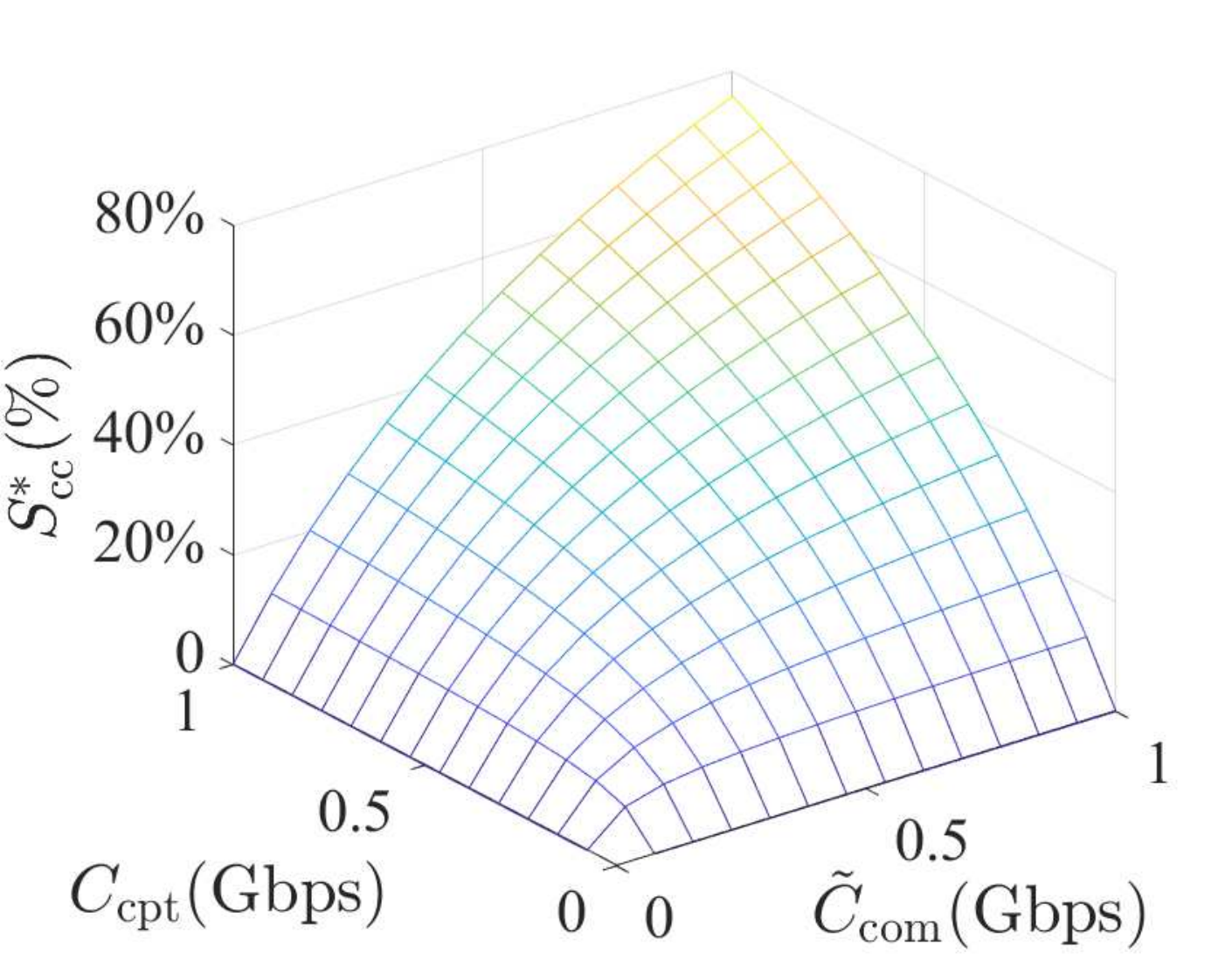}
        \end{minipage}
    }\\ \vspace{-2mm}
    \centering
    \subfloat[$T_{\mathrm{cc}}\in(T_{\mathrm{seg}},2T_{\mathrm{seg}}) (T_{\mathrm{cc}}=1.5$ s) ]{\label{Fig:UnsetReg}
        \begin{minipage}[c]{0.7\linewidth}
            \centering
            \includegraphics[width=1\textwidth]{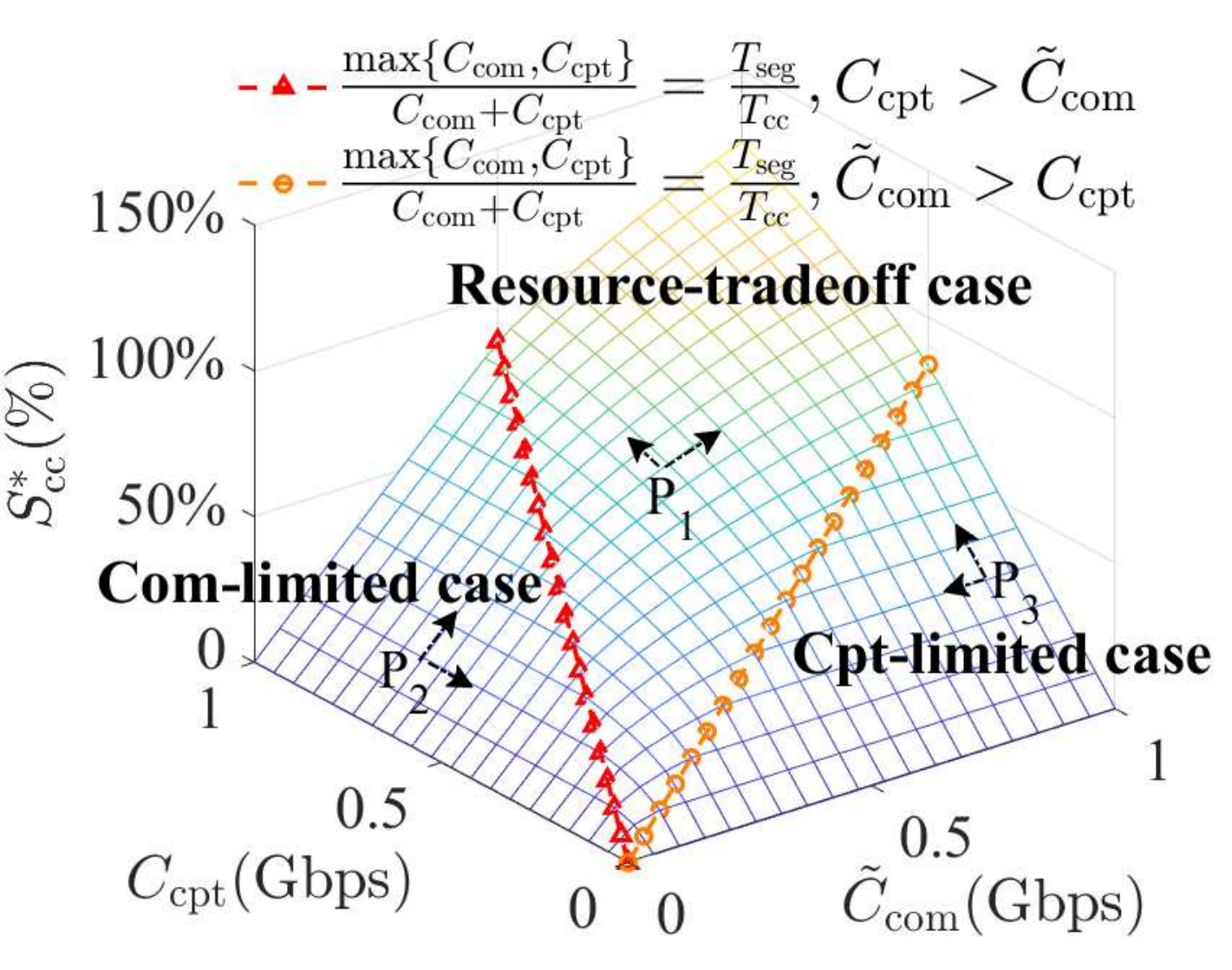}
        \end{minipage}
    }
    \vspace{-0.1cm}
    \caption{(a) Minimum-resource-limited region, (b) Unconditional resource-tradeoff region, (c) Conditional resource-tradeoff region.}\label{Fig:Regions}	
    \vspace{-0.36cm}
\end{figure}

In Fig. \ref{Fig:Regions}, we illustrate the three regions.
As shown in Fig. \ref{Fig:MinResReg}, if $C_{\mathrm{com}} \neq C_{\mathrm{cpt}}$, then the system performance is restricted either by communication or computing resource. By contrast, in the unconditional resource-tradeoff region shown in Fig. \ref{Fig:ResMatReg}, the communication and computing resources can be flexibly adjusted.
In the conditional resource-tradeoff region in Fig. \ref{Fig:UnsetReg}, the system configured with different resources lies in communication-limited case, resource-tradeoff case, or computing-limited case. The boundary of the three cases is $\frac{\max\{C_{\mathrm{com}},C_{\mathrm{cpt}}\}}{C_{\mathrm{com}} + C_{\mathrm{cpt}}}= \frac{T_{\mathrm{seg}}}{T_{\mathrm{cc}}}$. We can observe that if the system is resource-limited, say $P_3$ in the figure, no matter if we increase the computing rate or reduce the transmission rate in order to satisfy the condition for efficient resource configuration (i.e., $\frac{\max\{C_{\mathrm{com}},C_{\mathrm{cpt}}\}}{C_{\mathrm{com}} + C_{\mathrm{cpt}}}\leq \frac{T_{\mathrm{seg}}}{T_{\mathrm{cc}}}$), the system will finally fall into the resource-tradeoff case.

\begin{figure}[htbp]
\vspace{-0.5cm}
    \centering
    \subfloat[$T_{\mathrm{cc}}< T_{\mathrm{seg}} (T_{\mathrm{cc}}=0.9$ s)]{\label{Fig:ConVio_1}
        \begin{minipage}[c]{0.5\linewidth}
            \centering
            \includegraphics[width=1\textwidth]{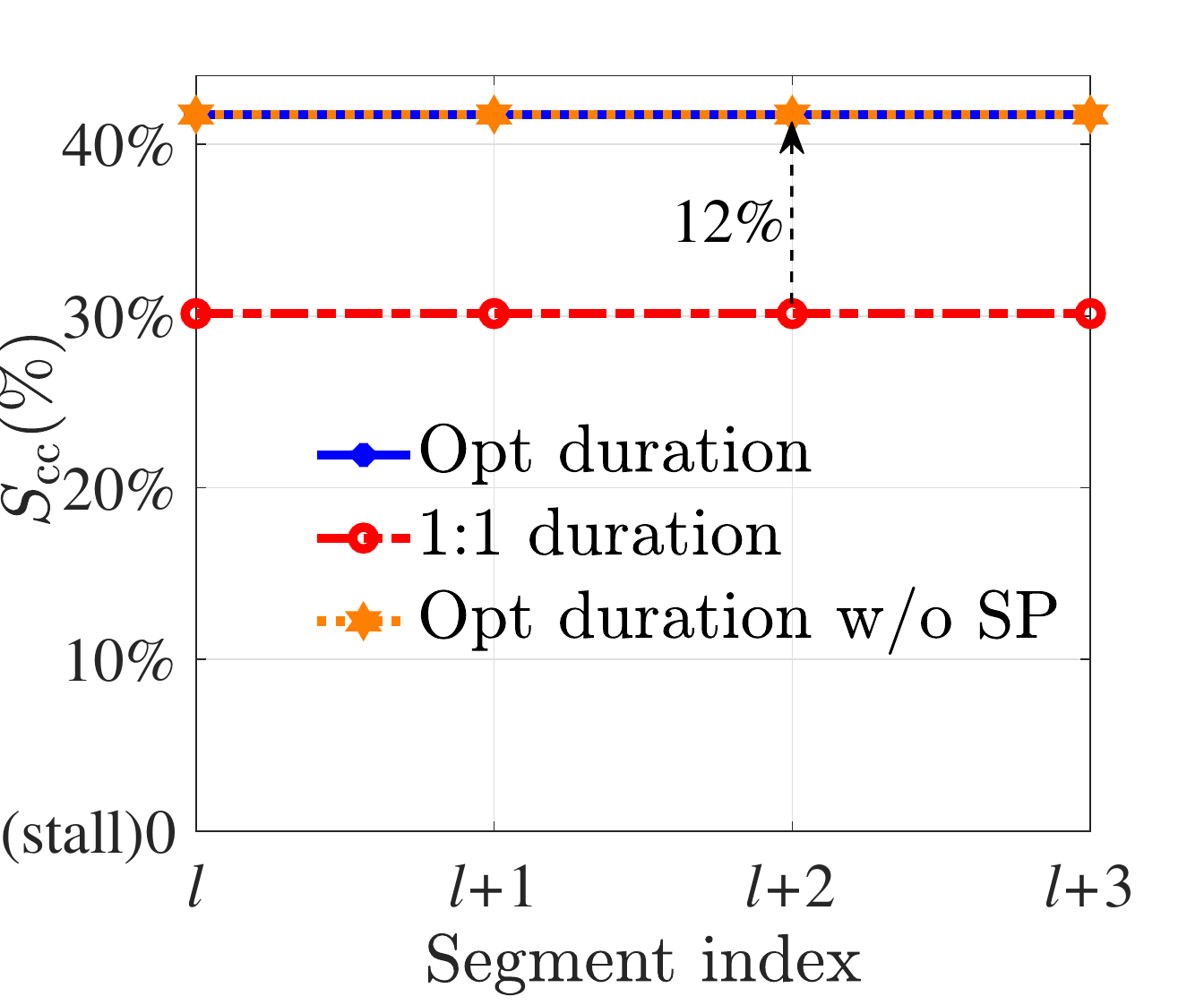}
        \end{minipage}
    }
    \subfloat[$T_{\mathrm{cc}}\in( T_{\mathrm{seg}},2T_{\mathrm{seg}}) (T_{\mathrm{cc}}=1.5$ s)]{\label{Fig:ConVio_2}
        \begin{minipage}[c]{0.5\linewidth}
            \centering
            \includegraphics[width=1\textwidth]{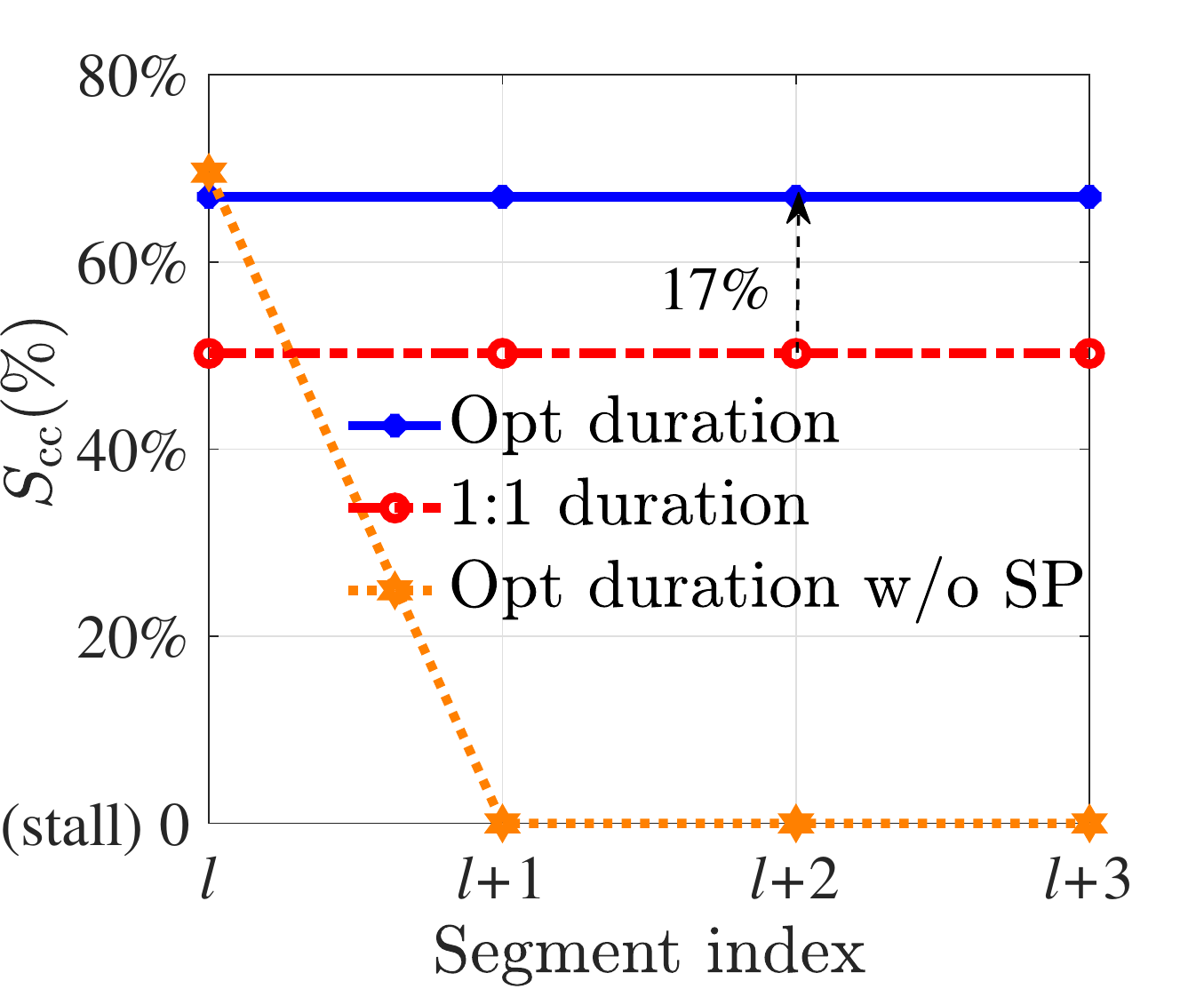}
        \end{minipage}
    }\\
    \subfloat[$T_{\mathrm{cc}}\in(T_{\mathrm{seg}},2T_{\mathrm{seg}}) (T_{\mathrm{cc}}=1.5$ s) ]{\label{Fig:ConVio_3}
        \begin{minipage}[c]{0.5\linewidth}
            \centering
            \includegraphics[width=1\textwidth]{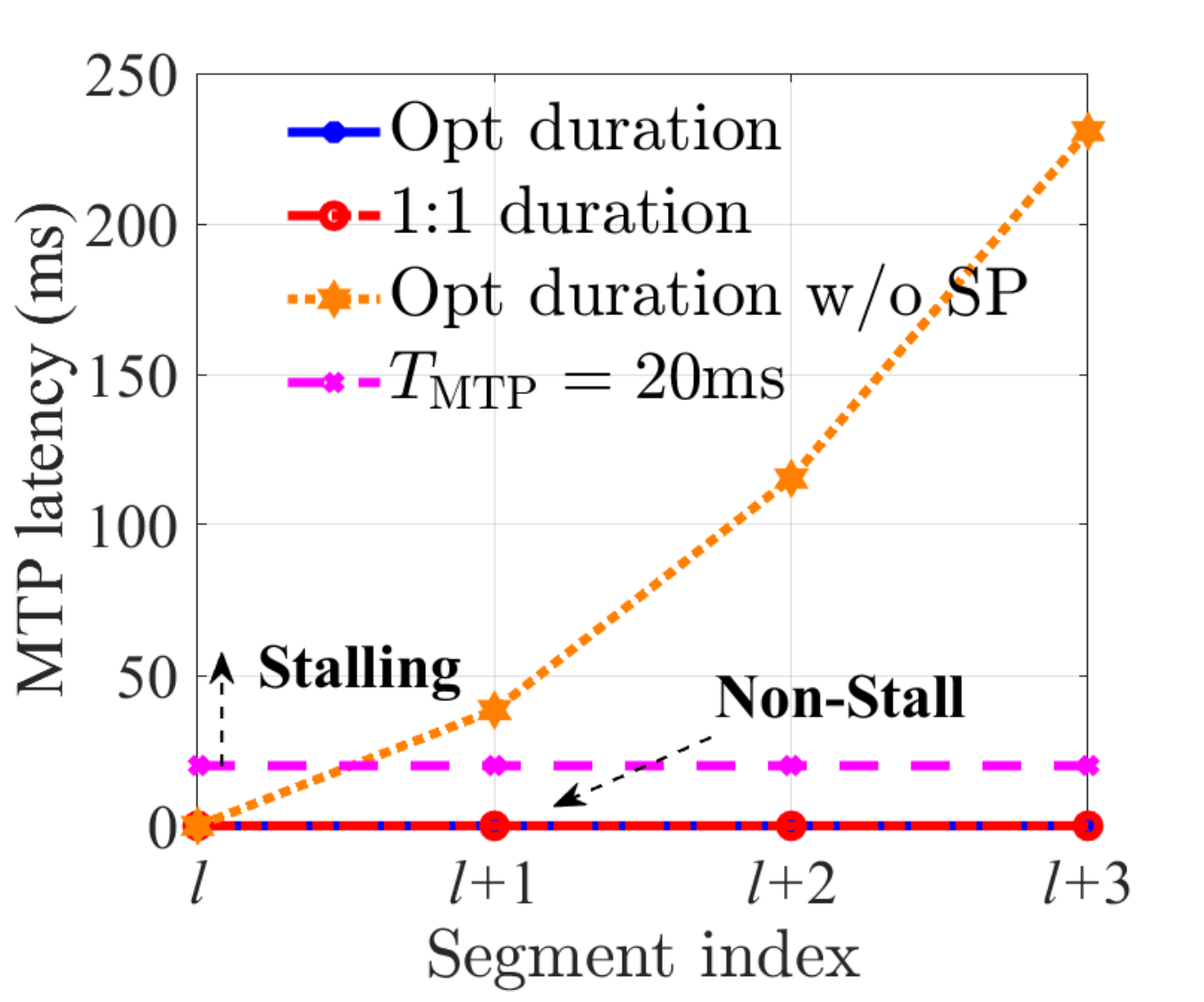}
        \end{minipage}
    }
    \subfloat[$T_{\mathrm{cc}}>2 T_{\mathrm{seg}} (T_{\mathrm{cc}}=2.1$ s)]{\label{Fig:ConVio_4}
        \begin{minipage}[c]{0.5\linewidth}
            \centering
            \includegraphics[width=1\textwidth]{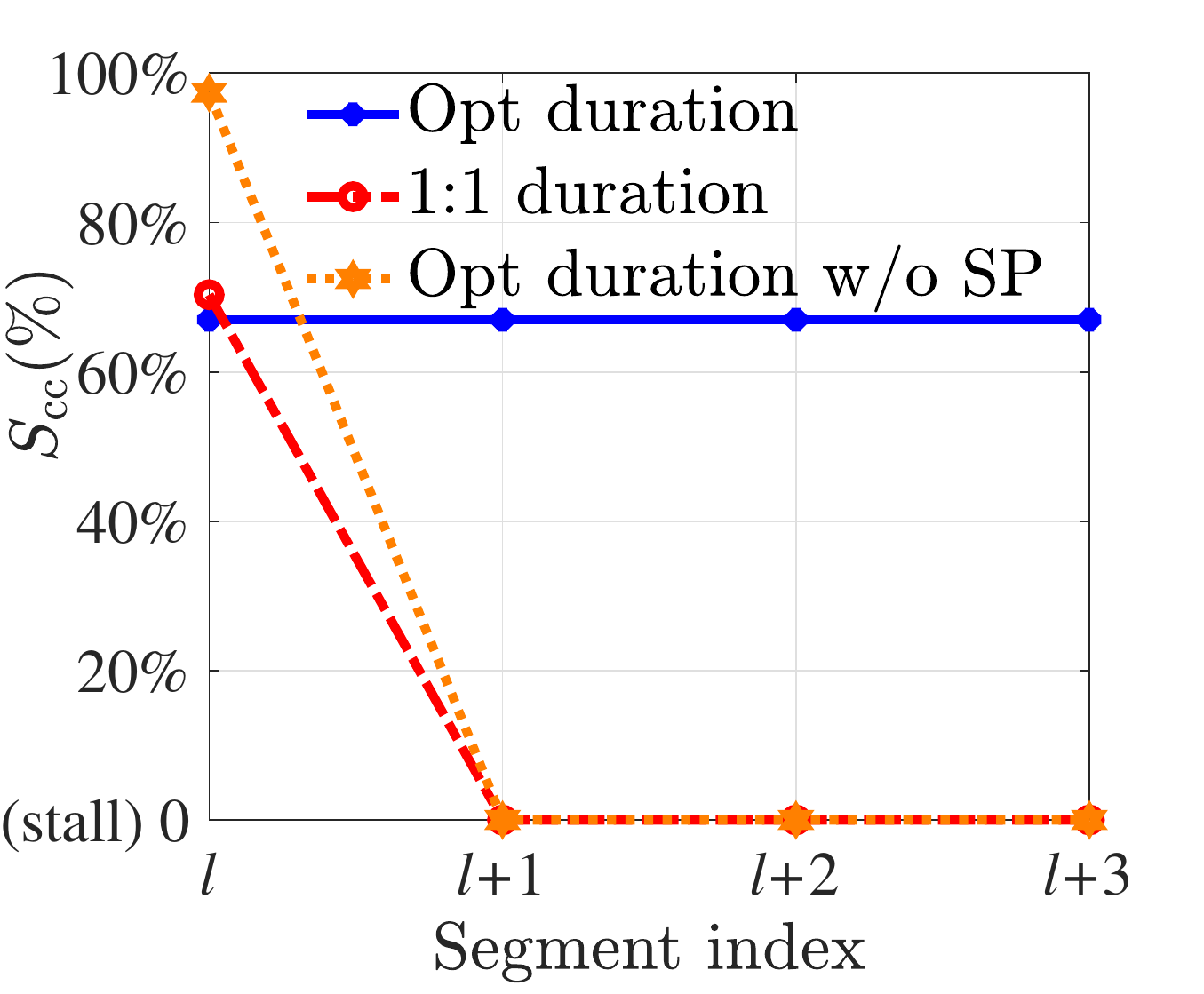}
        \end{minipage}
    }
    \vspace{-0.15cm}
    \caption{$S_{\mathrm{cc}}$ and MTP latency v.s. segment index, $\tilde{C}_{\mathrm{com}}=900$ Mbps and $C_{\mathrm{cpt}}=400$ Mbps.}\label{Fig:ConVio}	
    \vspace{-0.6cm}
\end{figure}

In Fig. \ref{Fig:ConVio}, we verify the necessity of imposing the duration-squeezing-prohibited constraints by taking the value of $S_{\mathrm{cc}}$ over the first four proactively streamed segments
as an example (the results for other values of $\tilde{C}_{\mathrm{com}}$ and $C_{\mathrm{cpt}}$ are similar whenever the difference between the two values are more than 500). We compare the optimal durations in \eqref{opt_slt} with two baseline schemes without considering the duration-squeezing-prohibited (SP) constraints. One is the optimal solution of problem \textbf{P1} without the SP constraints in \eqref{queue_stable_cpt} and \eqref{queue_stable_com}, where $t_{\mathrm{com}}=t_{\mathrm{com}}^o$ and $t_{\mathrm{cpt}}=t_{\mathrm{cpt}}^o$, with legend ``opt duration w/o SP". The other scheme fixes the durations as $t_{\mathrm{com}}=\frac{T_{\mathrm{cc}}}{2}$, with legend ``1:1 duration".
As expected, the optimal durations yield the best performance from the ($l$+1)th segment.

When $T_{\mathrm{cc}}<T_{\mathrm{seg}}$ as shown in Fig. \ref{Fig:ConVio_1}, the optimal durations achieve the same performance as the baseline ``opt duration w/o SP", because $T_{\mathrm{cc}}\leq T_{\mathrm{seg}}$ is the sufficient condition of \emph{Case 2}. When \emph{Case 2} holds, $t_{\mathrm{com}}^o,t_{\mathrm{cpt}}^o\leq T_{\mathrm{seg}}$, i.e., the transmitting and computing with ``opt duration w/o SP" will not cause the squeeze. These two schemes outperform the scheme ``1:1 duration", which shows the gain of matching the imbalanced computing rate and transmission rate.

When $T_{\mathrm{seg}}<T_{\mathrm{cc}}<2T_{\mathrm{seg}}$ as shown in Fig. \ref{Fig:ConVio_2}, although ``opt duration w/o SP" slightly outperforms the optimal durations for the $l$th segment, the completion rate of the CC tasks of this baseline degrades to zero and stalling happens for the ($l$+1)th segment. This is because $T_{\mathrm{c}}^{\max}=\frac{\max\{\tilde{C}_{\mathrm{com}},C_{\mathrm{cpt}}\}T_{\mathrm{cc}}}{\tilde{C}_{\mathrm{com}} + C_{\mathrm{cpt}}}=1.04>T_{\mathrm{seg}}$, i.e., \emph{Case 1} holds, where either the transmitting or the computing of this baseline for the $l$th segment squeezes the duration for the ($l$+1)th segment that causes the playback stalling, as visualized in Fig. \ref{Fig:ConVio_3}. For the three schemes, the motion-to-photon (MTP) latency of ($l$+$n$)th segment can be expressed as
$T_{\mathrm{MTP}}= \left[t_{\mathrm{com}} + t_{\mathrm{cpt}} - (n-1)\left((\Delta p)^+ + (\Delta m)^+\right)\right]^+$.

When $T_{\mathrm{cc}}>2T_{\mathrm{seg}}$ as shown in Fig. \ref{Fig:ConVio_4}, the squeeze is unavoidable for two baselines. This shows the necessity of imposing the duration-squeezing-prohibited constraints.


\vspace{-0.2cm}\section{Conclusion}\label{Conclusion}\vspace{-0.15cm}
In this paper, we investigated maximizing the completion rate of CC tasks with task duration-squeezing-aware constraint in proactive VR streaming. From the obtained closed-form solution,
we found the minimum-resource-limited, unconditional, and conditional resource-tradeoff regions. The boundary of the three regions depends on the relation between the total time budget for proactive communication and computing and the playback duration of a segment. In the minimum-resource-limited region, communication and computing resources can not be traded off.
In the unconditional resource-tradeoff region, the resources can be flexibly configured while in the conditional resource-tradeoff region, the efficient configuration should satisfy a condition. Numerical results validated the necessity of imposing duration-squeezing-prohibited constraints and illustrated these regions.

\bibliographystyle{IEEEtran}
\bibliography{IEEEabrv,ref}

\end{document}